# Quantum tunnelling in hydrogen atom transfer brings uncertainty to polymer degradation

Yu Zhu [1], Xinrui Yang [1], Famin Yu [1], Rui Wang [1], Qiang Chen [2],

Zhanwen Zhang [2,*] and Zhigang Wang [1,*]

[1] *Institute of Atomic and Molecular Physics, Jilin University, Changchun 130012, China*
[2] *Laser Fusion Research Center, China Academy of Engineering Physics, Mianyang 621900, China*
[*] E-mail: wangzg@jlu.edu.cn (Z. W.); bjzzw1973@163.com (Z. Z.)

## Abstract

The low degradability of common polymers composed of light elements, results in a serious impact on the environment, which has become an urgent problem to be solved. As the reverse process of monomer polymerization, what deviates degradation from the idealized sequential depolymerization process, thereby bringing strange degradation products or even hindering further degradation? This is a key issue at the atomic level that must be addressed. Herein, we reveal that hydrogen atom transfer (HAT) during degradation, which is usually attributed to the thermal effect, unexpectedly exhibits a strong high-temperature tunnelling effect. This gives a possible answer to the above question. High-precision first-principles calculations show that, in various possible HAT pathways, lower energy barrier and stronger tunnelling effect make the HAT reaction related to the active end of the polymer occur more easily. In particular, although the energy barrier of the HAT reaction is only of $10^{-2}$ magnitude different from depolymerization, the tunnelling probability of the former can be 14~32 orders of magnitude greater than that of the latter. Furthermore, chain scission following HAT will lead to a variety of products other than monomers. Our work highlights that quantum tunnelling may be an important source of uncertainty in degradation and will provide a direction for regulating the polymer degradation process.

## Introduction

Degradation of widely used polymers containing light elements, has always been the focus of research owing to increasing environmental pollution (1-3). Compared with the monomer polymerization process, which can synthesize polymers with controlled structures and specific functions (4-6), degradation is much more complicated whose products are not just the monomer (7, 8). Moreover, it will produce many micron or nanoscale residues, thereby causing potential harm to human health and the ecological environment (9-14). These indicate that although degradation and polymerization are reversible at the atomic level, degradation may be affected by different effects, which should be taken seriously.

Notably, researchers have realized that hydrogen atom transfer (HAT) is a universal reaction during degradation, which will lead to changes in the polymer structure, thus may bring uncertainties in degradation. It has been well accepted that H atom on the main chain of polymers composed of light elements, such as carbon and hydrogen, can be seized by free radical to form a new intrachain radical which will cause chain scission (15, 16). Although experimental technology (e.g., in-situ measurements of spectra) has been greatly improved to study the reaction mechanism (17), direct experimental descriptions especially those at the atomic-level spatial resolution of this reaction are still lacking. Moreover, for such a long-distance HAT process that involves breaking the chemical bond or even overcoming the steric hindrance effect, it is not relatively easy to achieve from the perspective of classical thermal disturbance. Therefore, it is necessary to conduct research on HAT reactions in polymers at the atomic level based on a new perspective.

Actually, owing to the light mass of the H atom, many processes involving it are susceptible to quantum-mechanical effects, such as tunnelling (18-22). Tunnelling by H species is of critical importance for many reactions, such as proton-coupled electron transfer reactions (23, 24), redox enzyme reactions (25, 26), conformational inversion of molecules (27-30), and rotation of hydrogen bonds (31, 32). Generally, tunnelling is significant at low temperatures, although H tunnelling is not restricted to such temperatures (19, 28, 29). These attracted our attention: for polymers that usually degrade at room temperature or even higher temperatures, can the tunnelling effect in their HAT play a leading role under this temperature condition?

In this work, we focus on the HAT reactions to study their classical over-barrier process and through-barrier quantum tunnelling process based on the high-precision quantum mechanics calculations. Surprisingly, the results show that the strong high-temperature tunnelling effect presented in HAT is significant for degradation, which may be an important source of the complexity for degradation problems. These findings establish a new perspective for understanding HAT in polymer degradation, and will further provide references for regulating the degradation processes of polymers.

## Methods

To carry out this study, density functional theory (DFT) method of first-principles combined with the tunnelling probability calculation method were used to investigate the reaction pathways. A typical kind of common polymer containing light elements, poly-α-methylstyrene (PAMS), was selected as the calculation model due to the representative functional groups (benzene ring, methyl and hydrogen) on its side chain. Considering the reliability and resource requirements of quantum calculations, the di-radical PAMS tetramer was used, for which the rationality of the structure selection has been confirmed in our previous studies (33, 34). All structures (including reactants, transition states (TSs) and products) were fully optimized at the B3LYP-D3 (35, 36) level in conjunction with 6-31+G(d,p) basis sets. B3LYP is a commonly used functional for studying di-radical systems, which can produce reasonable energetic results for di-radical intermediates and transition states (37-39). Based on the optimized geometries, the nature of extreme points was evaluated by performing harmonic vibrational frequency calculations. The numbers of imaginary frequencies were used to confirm that the structure was a stable point or TS. Meanwhile, the zero-point energy (ZPE) scaled by the ZPE scaling factor was also determined. The IRC (40, 41) was calculated to ensure the reliability of the reaction process. Due to the existence of spin contamination in the low-spin state, we corrected it with an approximate spin-projection method (see Supplementary Information (SI) Part1). The above calculations were all performed using the Gaussian 09 package (42). For the dynamic simulation, the DFTB+19.1 package (43) was used based on the empirical dispersion corrected density functional tight-binding (DFTB-D) method (44, 45). A detailed description is given in the SI Part2.

For tunnelling probability ($P_{tunnelling}$), under the premise of the Born-Oppenheimer approximation, they are calculated with WKB approximation (46-48), as following formula: $P_{tunneling} = Exp\left[-\frac{2}{\hbar}\int_{x_1}^{x_2}\sqrt{2m(V(x)-E_p)}\,dx\right]$. Among them, m is the reduced mass on the vibration mode corresponding to the imaginary frequency of the TS. This vibration mode contains the main degrees of freedom involved in HAT or depolymerization. $V(x)$ is the potential energy function, which is obtained according to the IRC approach (40, 41). As the mass-weighted steepest descent path connecting reactants and products on the potential energy surface (PES), IRC has been widely used in the analysis and prediction of mechanisms for a variety of chemical reactions (49, 50). $E_p$ represents the provided energy. For the thermodynamic model, thermal disturbance probability ($P_{thermal}$) suits the Boltzmann distribution, as following formula: $P_{therm} = Exp(-\Delta E/k_b T)$, where $\Delta E$ is the difference from $E_p$ to the energy barrier.

## Results and discussion

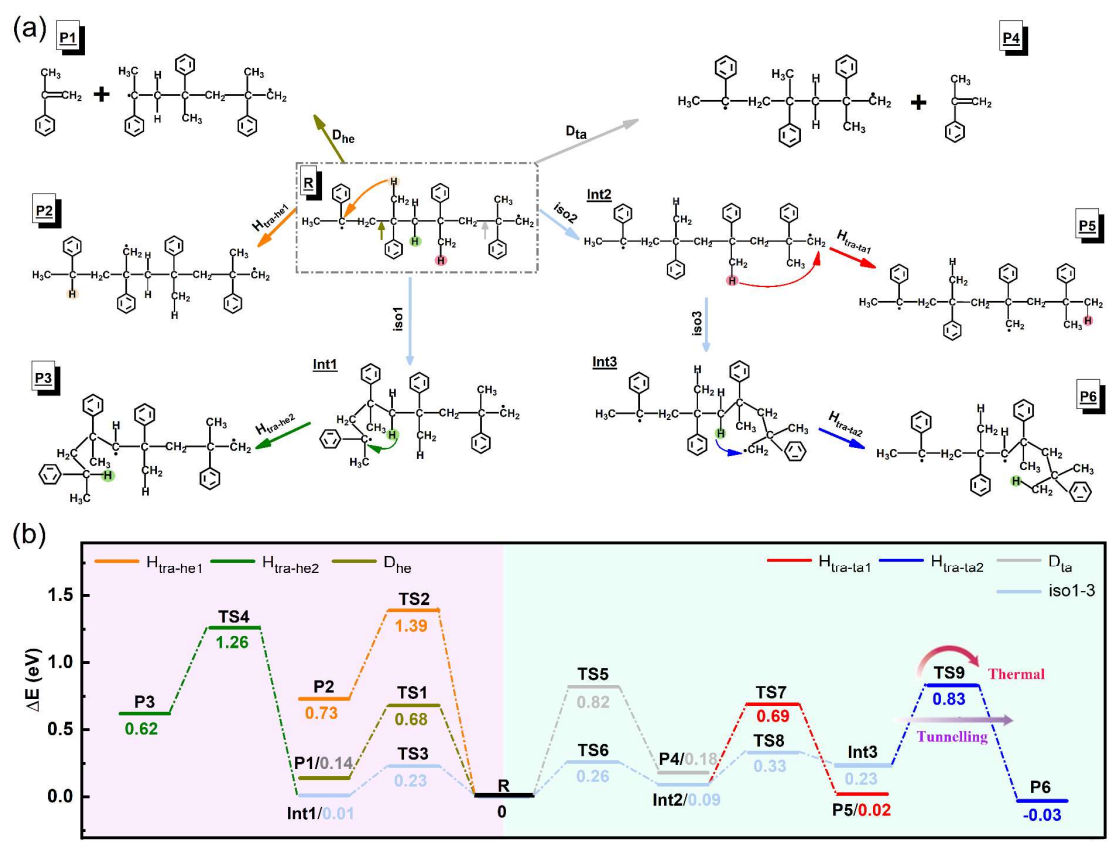

**Fig. 1 | Reactions pathways of PAMS. a**, Schematic diagram of reaction pathways. $H_{tra-he1}$ indicates that the H atom on the methyl group is transferred to the head end (abbreviated as 'he', C-unsaturated end). iso1-3 refer to isomerization processes. $H_{tra-he2}$ represents that the H atom on the methylene is transferred to the head end after occurring iso1. $H_{tra-ta1}$ indicates that the H atom on the methyl is transferred to the tail end (abbreviated as 'ta', $CH_2$-unsaturated end) after occurring iso2. $H_{tra-ta2}$ represents that the H atom on the methylene is transferred to the tail end after Int2 occurring iso3. $D_{he}$ and $D_{ta}$ are pathways that produce the AMS monomer, which have been studied in our previous work (33). **b**, Potential energy surfaces of these nine reactions. 'R', 'Int' and 'P' represent the reactant, intermediate and product, respectively. The energy of the reactant is taken as the zero. The pink and light blue areas in the figure represent the reaction pathways related to the head end and tail end, respectively. The energy values were all corrected by ZPE.

First, we obtained the possible reaction pathways by means of TS optimization and intrinsic reaction coordinate (IRC) analysis. Their schematic diagram and corresponding potential energy surfaces (PESs) are shown in Fig. 1.

For the di-radical structure, there are two kinds of HAT pathways at each head end (C-unsaturated end) and tail end (CH$_2$-unsaturated end). By taking the tail end as an example, due to the steric hindrance of CH$_2$ in the ortho position, the C atom at the tail end will seize the H atom on the methyl or methylene group of the adjacent monomer. However, because the nearby benzene ring also has steric hindrance, neither reaction is a simple one-step process. At first, this benzene ring will flip to form an intermediate (iso2). There are two cases afterwards. In the first case, the C atom at the tail end will directly seize the H atom from the neighbouring methyl group (H$_{tra-ta1}$). In the second case, a 'back-biting' process will continue to occur at the tail end, resulting in the formation of a 6-membered ring intermediate (iso3). Subsequently, the H atom from the methylene group on the second monomer (head-end monomer is the first) will be transferred to the tail end (H$_{tra-ta2}$). These two kinds of HAT reactions were also observed in our dynamic simulations (for details, see SI part3). Similarly, there are such two reactions at the head end. The difference is that the H atom on the methyl group of the second monomer can be transferred directly to the head end (H$_{tra-he1}$), whereas the transfer of the H atom on the methylene (H$_{tra-he2}$) requires a prerequisite: the polymer must first undergo an isomerization process to form a 5-membered ring intermediate (iso1).

Next, we analysed these reactions from the perspective of energy (Fig. 1(b)). For three isomerization processes, including the benzene ring flip, 'back-biting' of the tail end to form a 6-membered ring and 'back-biting' of the head end to form a 5-membered ring intermediate, their energy barriers are 0.26 eV, 0.24 eV and 0.23 eV, respectively. This indicates that the isomerization of PAMS easily occurs during degradation, especially the formation of polycyclic intermediates. In the two HAT reactions at the tail end, although the energy barriers for H$_{tra-ta1}$ and H$_{tra-ta2}$ are both 0.60 eV, considering that degradation is a reversible reaction, the relative energies of the product and TS (called the reverse energy barrier) also affect the occurrence of the reaction. The greater reverse energy barrier of H$_{tra-ta2}$ (0.86 eV) than that of H$_{tra-ta1}$ (0.67 eV) suggests that H atoms on the methylene are more likely to transfer, reflecting the selectivity of HAT reaction. This rule also applies to H$_{tra-he1}$ and H$_{tra-he2}$, for which the corresponding

**6**

energy barriers of these two reactions are 1.39 eV and 1.25 eV. Additionally, the HAT reaction is more likely to occur at the tail end. In a previous study (33), we have investigated the depolymerization reactions of PAMS. The energy barriers required for the dissociation of the head and tail end to generate monomer are $D_{he}$ (0.68 eV) and $D_{ta}$ (0.82 eV), respectively. By comparison, it was found that the tail end is more prone to occur HAT rather than depolymerization.

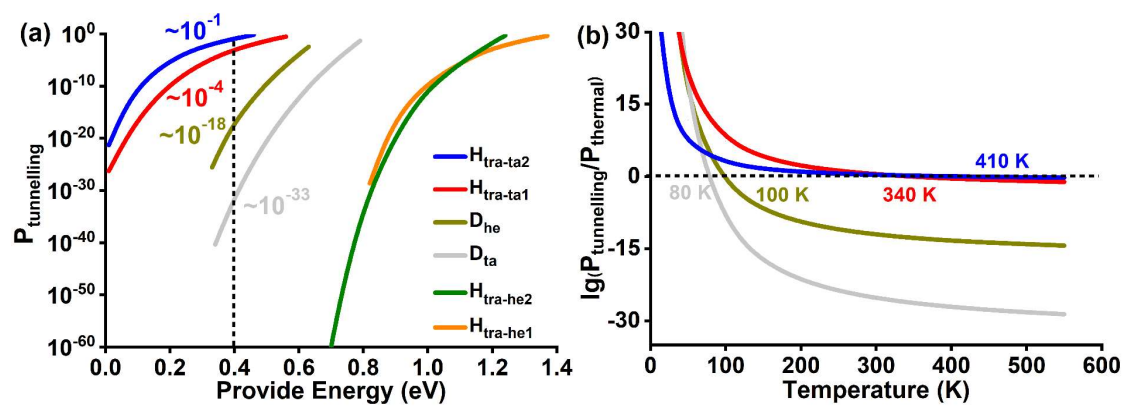

**Fig. 2 | Probabilities of quantum tunnelling ($P_{tunnelling}$) and thermal disturbance ($P_{thermal}$) of reaction pathways. a**, $P_{tunnelling}$ as a function of the provided energy ($E_p$). "~" denotes the magnitude. **b**, Ratio of $P_{tunnelling}$ to $P_{thermal}$ of reaction pathways when $E_p$ is 0.40 eV. The black dotted line means that the two kinds of probabilities are equal.

In fact, polymers can not only achieve HAT by overcoming the energy barrier, but also instantaneous H transfer through quantum tunnelling. Using different excitation energies, we calculated the probabilities of quantum tunnelling ($P_{tunnelling}$) for PAMS to achieve HAT and monomer dissociation based on the Wenzel-Kramers-Brillouin (WKB) approximation (46-48). Here, the provided energy ($E_p$) should be higher than the energies of the reactant and product in the reaction, thus giving $P_{tunnelling}$, as shown in Fig. 2a. With the increase of $E_p$, $P_{tunnelling}$ increases gradually. When the same excited energy is provided, the $P_{tunnelling}$ of tail-end HAT ($H_{tra-ta1}$ and $H_{tra-ta2}$), depolymerization ($D_{he}$ and $D_{ta}$) and head-end HAT ($H_{tra-he1}$ and $H_{tra-he2}$) decreases sequentially. The magnitudes of $P_{tunnelling}$ for $H_{tra-he1}$ and $H_{tra-he2}$ are similar. Considering that the energies of the products for these two reactions are relatively large, the required $E_p$ is also high, thus, we will not discuss their $P_{tunnelling}$ in detail later. Moreover, when the $E_p$ is 0.4 eV, the magnitude of $P_{tunnelling}$ for $H_{tra-ta1}$ and $H_{tra-ta2}$ are $10^{-1}$ and $10^{-4}$, and those for $D_{he}$ and $D_{ta}$ are $10^{-18}$ and $10^{-33}$, showing obvious differences. Therefore, compared with the dissociation of the monomer, the HAT reaction at the tail end has a strong tunnelling effect, which can easily occur through tunnelling.

Meanwhile, from the classical perspective, the thermal effect will cause the reaction to occur with a certain probability. We further compared the thermal disturbance probability ($P_{thermal}$) and $P_{tunnelling}$ by calculating the ratios of the two at different temperatures and $E_p$. Here, to clearly see the difference between $P_{thermal}$ and $P_{tunnelling}$, we ignore the possible correlation between the two, which has also been implemented in many researches (51, 52). Fig. 2b shows the ratios when the $E_p$ is 0.40 eV. It can be found that the temperatures at which tunnelling dominates in $H_{tra-ta2}$, $H_{tra-ta1}$, $D_{he}$ and $D_{ta}$ are approximately 410 K, 340 K, 100 K and 80 K. This suggests that H tunnelling during polymer degradation can play a significant role even at room temperature. Similar laws are also obtained under other $E_p$ (details are provided in SI part 4). Actually, it is now well established that H tunnelling contributions can be substantial at room temperature (19, 22, 53), but this effect was not previously noticed in polymer degradation. The main role of quantum tunnelling in the HAT reaction indicates that the polymer may undergo frequent HAT even at temperatures where degradation does not occur. Moreover, it worth mentioning that the occurrence of HAT will lead to the change of the polymer structure, resulting in the uncertainty of degradation. Therefore, it can be reasonably speculated that the HAT reaction promoted by quantum tunnelling may be an important source of uncertainty in degradation.

Furthermore, we study the chain scission pathways after HAT from the perspective of PESs. Also taking $H_{tra-ta1}$ as an example, there are two possible scission pathways after HAT. In the first pathway, a monomer free radical and a chain with a C-unsaturated end are produced. In the second pathway, a dimer and an unsaturated chain form together. The chain scission reactions that occur after the other three HAT reactions also produced various products other than monomers, such as monomer-like, dimer-like and trimer-like molecules (details can be seen SI part5). The analyses of PESs further confirm that HAT contributes to the occurrence of chain scission reactions, thus resulting in the uncertainty of degradation.

Classical kinetic or thermodynamic control is conventionally considered an important way to understand and predict the diversity of chemical reactions. As an increasing number of tunnelling phenomena in chemical reactions were revealed, researchers have gradually realized that tunnelling control may be regarded as a type of nonclassical kinetic control (20, 28). Moreover, an experimental study of double-hydrogen tunnelling in porphycene molecules based on scanning tunnelling microscope (STM) found that the height of the STM tip affects the tunnelling properties (30). Thus, it is conceivable that in future experiments, tunnelling can be manipulated through external stimuli, which provides a new idea for regulating the polymer degradation process.

## Conclusion

In summary, our study revealed the important effect of H atom tunnelling on polymer degradation. The calculation of PESs shows that there are two possible HAT pathways at each end of the main chain, namely, the H atom on the methyl or methylene of ortho monomer transfers to the head end and tail end. Among them, for two reactions at the same chain end, the H atom on methylene is more likely to transfer, which reflects the selectivity of the HAT reaction. While, by comparing the energy barriers of reactions at different chain ends, it can be found that HAT occurs more easily at the tail end. More importantly, although the energy barriers required for tail-end HAT and depolymerization differ slightly, their tunnelling probabilities are vastly different. At a provided energy of 0.4 eV, the tunnelling probability of the former is 14-32 orders of magnitude greater than that of the latter. In particular, the tunnelling effect of the former can still dominate at room temperature. Thus, it is known that lower energy barrier and stronger high-temperature tunnelling effect result in a reaction in which H atom transfer to the tail end occur more easily. Additionally, HAT will further promote the occurrence of chain scission to form products other than monomers, thus bringing about the uncertainty of degradation. The above results reveal that tunnelling is an important source of the complexity of polymer degradation, which will provide a guide to better control the polymer degradation process in the future.

## Acknowledgements

This work was supported by the National Natural Science Foundation of China (grant numbers 11974136 and 11674123). Z. Wang also acknowledges the assistance of the High-Performance Computing Center of Jilin University and National Supercomputing Center in Shanghai.

## Author contributions

Y. Zhu performed the theoretical simulations, Z. Wang initiated and designed the work. Z. Wang and Z. Zhang supervised the work. Y. Zhu, X. Yang, F. Yu, R. Wang, Q. Chen, Z. Zhang and Z. Wang discussed the results. Y. Zhu, Z. Zhang and Z. Wang wrote the article.